\documentclass[5p,preprint]{elsarticle}
\biboptions{sort&compress}
\usepackage{lineno,isotope,mathrsfs}\modulolinenumbers[5]
\usepackage{amsmath,mathpazo,mathptmx,epsfig,bm,mathrsfs,feynmp}
\usepackage{color,slashed,nicefrac,exscale,multirow,soul}
\usepackage{times,txfonts,xcolor,graphicx,gensymb,tikz}
\usepackage[normalem]{ulem}
\usepackage[breaklinks,colorlinks,citecolor=blue]{hyperref}
\definecolor{red}{rgb}{0.8,0,0}
\definecolor{violet}{rgb}{0.4,0,0.4}
\definecolor{green}{rgb}{0,0.5,0.0}
\definecolor{navy}{rgb}{0.0,0.0,0.6}
\definecolor{orange}{rgb}{0.8,0.2,0.0}

\usepackage[normalem]{ulem} 

\newcommand{\bea}{\begin{eqnarray}}
\newcommand{\eea}{\end{eqnarray}}

\begin{document}
\begin{frontmatter}
\title{The hyperon superfluidity and the hyperon couplings in neutron stars within the relativistic mean field model}
\author[addr1,addr2]{Zhong-Hao Tu}
\ead{tuzhonghao@mail.itp.ac.cn}
\author[addr1,addr2,addr3]{Shan-Gui Zhou\corref{cor1}}
\ead{sgzhou@itp.ac.cn}
\address[addr1]{CAS Key Laboratory of Theoretical Physics, Institute of Theoretical Physics,
	Chinese Academy of Sciences, Beijing, 100190, China}
\address[addr2]{School of Physical Sciences, University of Chinese Academy of Sciences, Beijing, 100049, China}
\address[addr3]{Peng Huanwu Collaborative Center for Research and Education, Beihang University, Beijing, 100191, China}
\cortext[cor1]{Corresponding author}

\begin{abstract}
 A systematic study of the effects of hyperon couplings on hyperon superfluidity is conducted by using the relativistic mean field model. Combining the slope of symmetry energy, the hyperon couplings are determined in two ways---by the hypernuclear potentials or under the SU(3) symmetry. In either way, the hyperon coupling constants cannot be fixed uniquely but vary within a certain range due to the uncertainties in hypernuclear potentials or the breaking of SU(6) to SU(3) symmetry. When the coupling constants are constrained by the hypernuclear potentials, the pairings of $\Lambda$ and $\Xi^{0,-}$ are strong and they each show little variations. The pairing of $\Sigma^-$ is more sensitive to hyperon potentials and the slope of symmetry energy. Under the SU(3) symmetry, the superfluidity of various hyperons differ significantly. The dependence of the pairings of $\Lambda$ and $\Xi^{0,-}$ on the additional parameters of SU(3) symmetry are the opposite to that of the maximum mass of neutron stars on the additional parameters of SU(3) symmetry, while the pairing of $\Sigma^-$ shows a similar trend in general. These results suggest that the hyperon superfluidity associated with astrophysical processes is an essential window to probe the physics of neutron star cores, the hyperon-hyperon interactions and the SU(3) symmetry. Compared with other hyperons, $\Sigma^-$ could serve as a cleaner glass for this purpose.
\end{abstract}
\end{frontmatter}
\section{Introduction}
\label{sec:intro}
Our knowledge of the interior of neutron stars (NSs), especially in the region of the baryon number density $\rho_{B}$ higher than two times the saturation density $\rho_{0}$, is uncertain yet. Hyperons could appear in NS cores once the nucleon Fermi energy exceeds the (in-medium) rest masses of the hyperons \cite{Glendenning1985_ApJ293-470}. A number of works have been devoted to the study of NSs composed of hyperons \cite{Glendenning1985_ApJ293-470,Schaffner1996_PRC53-1416,Li2007_CP16-1934,Weissenborn2012_PRC85-065802,Oertel2015_JPG42-075202,Katayama2015_PLB747-43,Providencia2019_FASS6-13,Hong2019_CTP71-819,Thapa2021_PRD103-063004,Rather2021_ApJ917-46,Rong2021_PRC104-054321,Tu2022_ApJ925-16}.
However, when hyperons are included, the direct Urca (dUrca) processes involving hyperons significantly enhance the neutrino luminosity and rapidly cool NSs \cite{Prakash1992_ApJ390-L77,Pethick1992_RMP64-1133,Haensel1994_AA290-458,Prakash1994_PR242-297,Ji1998_PRD57-5963}, and finally result in the fact that the surface temperatures of NSs are much lower than those observed \cite{Schaab1998_ApJ504-L99,Xu2011_CTP56-521,Vidana2013_NPA914-367}. For fixing this serious problem, hyperons are postulated to form a superfluid to suppress neutrino emission and ensure that the cooling curve of NSs and the astrophysical observations are consistent.

Hyperons can be paired after they are populated inside NS cores \cite{Schaab1998_ApJ504-L99,Balberg1998_PRC57-409,Takatsuka2001_PTP105-179,Tanigawa2003_PRC68-015801,Vidana2004_PRC70-028802,Zhou2005_PRL95-051101}.
The paired hyperons exponentially suppress the neutrino emissivity of hyperon dUrca processes far below the critical temperature $T_{c}$ of hyperon pairing \cite{Yakovlev2001_PR354-1,Yakovlev2004_ARAA42-169,Page2006_ARNPS56-327} and affect the dominant neutrino emission mechanism, i.e., the Cooper pair breaking and formation (PBF) mechanism, near $T_{c}$ \cite{Flowers1976_ApJ205-541,Yakovlev2001_PR354-1,Leinson2006_PLB638-114}. Several studies have indicated that the hyperon superfluidity plays an important role in NS cooling \cite{Page2004_ApJSupp155-623,Page2009_ApJ707-1131,Page2011_PRL106-081101,Newton2013_ApJL779-L4,Leinson2015_PLB741-87,Beznogov2016_MNRAS463-1307,Han2017_PRC96-035802,Raduta2017_MNRAS475-4347,Beloin2018_PRC97-015804,Dong2018_ApJ862-67,Fortin2018_MNRAS475-5010,Raduta2019_MNRAS487-2639,Sedrakian2019_PRD99-043011,Wei2020_MNRAS498-344,Bhat2020_EPJC80-544}.

The magnitude of the hyperon superfluidity in NSs is characterized by the hyperon-hyperon ($YY$) pairing gap. The uncertainties in determining the pairing gap arise from the ambiguous nuclear interactions above $\rho_0$ (including nucleon-nucleon ($NN$), nucleon-hyperon ($NY$), hyperon-hyperon, etc.) and $YY$ pairing force that is limited to theoretical aspect currently. We are mainly interested in the $^1S_0$ pairing gap inside NS cores in this work. Although the $YY$ pairing forces are still uncertain, several studies have given the values of the hyperon pairing gaps based on different available $YY$ pairing forces \cite{Vidana2004_PRC70-028802,Takatsuka2006_PTP115-355,Wang2010_PRC81-025801,Raduta2017_MNRAS475-4347,Sedrakian2019_EPJA55-167,Tu2022_PRC106-025806}.
But the study of the effects of the nuclear interactions on the hyperon superfluidity is still rare, especially for $NY$ and $YY$ interactions.

The nuclear interaction also determines the equation of state (EoS), interior composition and global properties of NSs. In this work, the relativistic mean field (RMF) model is utilized to calculate the EoSs of NSs. The nuclear interaction is characterized by the coupling constants between baryons and mesons. For an NS composed of necleons and hyperons, its EoS is certainly affected by the hyperon couplings,  specifically the coupling constants between hyperons and mesons in the RMF model. The main motivation of this work is to build the relation between the hyperon couplings and hyperon superfluidity by taking into account the uncertainties of nuclear interactions. Two approaches are employed for determining the hyperon couplings. One is the empirical hypernuclear potentials, e.g., $U_{\Sigma}^{(N)}$ and $U_{\Xi}^{(N)}$. These hypernuclear potentials are uncertain so far and hence we can use them to adjust the magnitude of the attractive interaction brought by the scalar mesons within their uncertainties \cite{Weissenborn2012_NPA881-62,Providencia2019_FASS6-13}. The other one is the SU(3) model. The SU(3) symmetry can provide a possibility of the existence of massive NS, we can relax the SU(6) symmetry to a more general SU(3) symmetry and determine the repulsive interactions originating from the couplings of the vector mesons and hyperons  \cite{Weissenborn2012_PRC85-065802,Weissenborn2013_NPA914-421,Lopes2014_PRC89-025805,Lim2018_PRD97-023010,Li2018_EPJA54-133,Fu2022_PLB834-137470}.
Combining with the indeterminate hypernuclear potentials and the SU(3) symmetry, we systematically study the effects of hyperon couplings on the hyperon superfluidity for the first time.

This paper is organized as follows. In Sec.~\ref{sec:theo},  the methodology for calculating the EoS of NSs and the pairing gaps of hyperons are given briefly. Sec.~\ref{sec:discussion_hyperU} shows how the empirical hypernuclear potentials affect the hyperon superfluidity. In Sec.~\ref{sec:discussion_SU3}, we discuss the effects of hyperon couplings on the hyperon superfluidity under the SU(3) symmetry. Finally, a brief summary and perspective are given in Sec.~\ref{sec:summary}.

\section{Theoretical Framework}
\label{sec:theo}
Using the well-known Bardeen-Cooper-Schrieffer (BCS) approximation \cite{Khodel1996_NPA598-390,Balberg1998_PRC57-409,Wang2010_PRC81-025801} and the separable pairing force \cite{Tian2009_PLB676-44,Rong2020_PLB807-135533}, the $^{1}S_0{}$ pairing gap $\Delta_{B}(k)$ of the baryon species $B$ at zero temperature is obtained by solving the gap equation
\begin{equation}\label{equ:gap_equ}
\begin{aligned}
    1 & = -\frac{1}{4\pi^{2}}\int k^{2}\mathrm{d}k\frac{G_{B}p^{2}(k)}{\sqrt{[E_{B}(k)-\mu]^{2}+\Delta_{0}^{2}p^{2}(k)}},
\end{aligned}
\end{equation}
where $E_{B}(k)$ is the single particle energy of the baryon species $B$, the chemical potential $\mu$ is determined by the value of $E_{B}(k)$ at the Fermi surface, i.e., $\mu = E_{B}(k_{\mathrm{F}})$ with the Fermi momentum $k_{\mathrm{F}}$. $G_{B}$ is the pairing strength of the baryon species $B$, $p(k)=\exp(-\alpha^2k^2)$ with the finite range $\alpha$ is the momentum dependence of pairing force. The separable pairing force of $BB$ pairing is $G_{B}(k,k')=-G_{B}p(k)p(k')=-G_{B}\exp(-\alpha^2k^2)\exp(-\alpha^2k'^2)$. $\Delta(k) = \Delta_0 p(k)$ is the trivial solution of the Eq.~(\ref{equ:gap_equ}) and we mainly refer the pairing gap at the Fermi surface $\Delta^{\mathrm{F}}=\Delta_0 p(k_{\mathrm{F}})$ in the present work. For $NN$ pairing, the pairing strength $G_{N}=738~\mathrm{MeV\cdot fm}^3$ and the finite range parameter $\alpha = 0.636~\mathrm{fm}$. A new set of hyperon pairing strengths have been proposed in the form of the separable pairing force \cite{Tu2022_PRC106-025806}. In this work, we choose $G_{\Lambda}=2G_{N}/3$ for $\Lambda$ hyperon, the upper limit of the $\Sigma\Sigma$ pairing strength $G_{\Sigma} = 2G_{N}/3$ for $\Sigma$ hyperon, and the lower limit of $\Xi\Xi$ pairing strength $G_{\Xi}=G_{N}$ for $\Xi$ hyperon~\cite{Tu2022_PRC106-025806}. The finite range of $YY$ pairings are the same as that of $NN$ pairing.

The EoS and the single particle energies of each baryons are obtained by using the RMF model. In this model, the $J^P=\frac{1}{2}^+$ baryonic octet interact with each other through the exchange of the scalar mesons ($\sigma$ and $\sigma^*$) and vector mesons ($\omega$, $\rho$ and $\phi$). The Lagrangian density which baryons interact with mesons is given by
\begin{equation}\label{equ:DDRMF_Lagrangian}
\begin{aligned}
    \mathcal{L}_{\mathrm{int}} = & \sum_{B}\bar{\psi}_{B}\left[ \gamma^{\mu}(-g_{\omega B }\omega_{\mu}
                 -g_{\rho B }\boldsymbol{\rho}_{\mu}\boldsymbol{\tau}_{B}-g_{\phi B }\phi_{\mu})\right. \\
                 &-\left.(-g_{\sigma B }\sigma-g_{\sigma^{*}B}\sigma^{*})\right]\psi_{B},
\end{aligned}
\end{equation}
where $\psi_{B}$ is the Dirac field of the baryon species $B$; $\boldsymbol{\tau}_{B}$ is the Pauli matrices for isospin of the baryon species $B$; $\omega_{\mu}$, $\boldsymbol{\rho}_{\mu}$, $\sigma^{*}$ and $\phi_{\mu}$ denote the quantum fields of mesons, respectively. In the density dependent (DD) parameterization, e.g., DD-ME2 \cite{Lalazissis2005_PRC71-024312}, the coupling constant between a baryon $B$ and a meson $m$ is density dependent and is parameterized by the relation $g_{mB}(\rho_{B})=g_{mB}(\rho_{0})f_{m}(\rho_{B})$ with
\begin{align}
  f_{m}(\rho_{B})&=a_{m}\frac{1+b_{m}(\rho_{B}/\rho_{0}+d_{m})^{2}}{1+c_{m}(\rho_{B}/\rho_{0}+e_{m})^{2}}, \quad m=\sigma~\mathrm{and}~\omega,\label{equ:DDfunc_s_w}\\
  f_{m}(\rho_{B})&=\mathrm{exp}[-a_{m}(\rho_{B}/\rho_{0}-1)], \quad m=\rho. \label{equ:DDfunc_r}
\end{align}
The single particle energy in RMF model is written as
\begin{align}
  E_{B}(k) = \sqrt{k^2+M_{B}^{*2}}+g_{\omega B}\omega+g_{\rho B}\tau_3^{B}\rho + g_{\phi B}\phi + \Sigma_{R}, \label{equ:spe}
\end{align}
where the meson fields have been replaced by the their expectation values after the mean field approximation. $\Sigma_{R}$ is the rearrangement term originating from the density dependence of the coupling constants. The effective mass $M_{B}^{*}$ is given by
\begin{align}
  M_{B}^{*} = M_{B}+U_{B}^{S}= M_{B}-g_{\sigma B}\sigma-g_{\sigma^* B}\sigma^*, \label{equ:effmass}
\end{align}
where $M_{B}$ is the rest mass of the baryon species $B$.

The $NN$ interaction at density higher than $2\rho_0$ remains unknown. Due to the onset and $\beta$ equilibrium conditions for hyperons \cite{Glendenning1985_ApJ293-470,Tu2022_ApJ925-16}, the properties of hyperonic matter are affected by the nucleonic matter and hence are influenced by the $NN$ interaction. The characteristic coefficients of nucleonic matter, e.g., $K_{\mathrm{sat}}$, $Q_{\mathrm{sat}}$, $E_{\mathrm{sym}}$ and $L_{\mathrm{sym}}$, dominated by $NN$ interaction could affect the behavior of EoS of NSs at different density ranges. The value of $L_{\mathrm{sym}}$ determines the intermediate-density properties of nucleonic matter and hence influences the radius of the canonical $1.4M_{\odot}$ NS and the NS mass that allows the dUrca processes \cite{Lopes2014_BJP44-774}. For considering the uncertainty of $NN$ interaction in this work, based on DD-ME2, we choose four $NN$ effective interactions with different $L_{\mathrm{sym}}$ by changing $g_{\rho N}$ and $a_{\rho}$ but fixing $E_{\mathrm{sym}}=27.09$ MeV at $\rho_{B}=0.11$ fm$^{-3}$ \cite{Li2019_PRC100-015809,Fu2022_PLB834-137470}: $L_{\mathrm{sym}}=30$ MeV, $g_{\rho N}=3.0314$, $a_{\rho}=0.8889$; original DD-ME2 with $L_{\mathrm{sym}}=51.26$ MeV; $L_{\mathrm{sym}}=80$ MeV, $g_{\rho N}=3.8892$, $a_{\rho}=0.2576$; $L_{\mathrm{sym}}=100$ MeV, $g_{\rho N}=4.1544$, $a_{\rho}=0.0905$. It should be noted that the changes in the properties of finite nuclei induced by this adjustment are less than 2\%, indicating that these effective interactions are acceptable for this work.

The $NY$ and $YY$ interactions can be determined by using SU(3) symmetry and several hypernuclear potentials. We adopt the SU(3) symmetric model \cite{deSwart1963_RMP35-916,Schaffner1994_AoP235-35,Weissenborn2012_PRC85-065802,Li2018_EPJA54-133,Lim2018_PRD97-023010,Fu2022_PLB834-137470} that contains three lightest quarks and assume the mixing angle $\theta_{V}$, which describes the mixing between singlet and octet vector mesons, is ideal mixing. The mixing angle from the quadratic mass formula for mesons, i.e., $\theta_{V}\approx40^{\circ}$ \cite{Zyla2020_PTEP2020-1}, is very close to the ideal mixing angle $\theta_{V}=35.3^{\circ}$, and thus it is reasonable that we keep the ideal mixing for the vector mesons in the present work. The additional two parameters, i.e., $z$ and $\alpha_{V}$, are used to adjust the coupling constants of the vector mesons \cite{Weissenborn2012_PRC85-065802,Lim2018_PRD97-023010}. The $z$, which lies in the range of $0\leq z \leq 2/\sqrt{6}$ for keeping the interaction due to $\omega$ exchange to be repulsive for all baryons, is the ratio of the singlet and octet coupling constants in the SU(3) quark model. The $\alpha_{V}$ is the weight factor for the contributions of symmetric and the antisymmetric couplings relative to each other in the SU(3) symmetric model and lies in the range of $0 \leq \alpha_{V} \leq1$. We take $z$ and $\alpha_{V}$ as two free parameters. The ratios of $g_{mY}$ to $g_{mN}$ for vector mesons can be written out by $z$ and $\alpha_{V}$, see Refs. \cite{Weissenborn2012_PRC85-065802,Lim2018_PRD97-023010}. The SU(6) quark model combines the flavor SU(3) and spin SU(2) symmetries and gives $z=1/\sqrt{6}$ and $\alpha_{V}=1$ \cite{Schaffner1993_PRL71-1328,Schaffner1994_AoP235-35}.

The scalar meson couplings $g_{\sigma Y}$ and $g_{\sigma^* Y}$ are determined by fitting empirical
hypernuclear potentials $U_{Y}^{(N)}$ and $U_{Y}^{(Y)}$ in symmetric nuclear matter at the saturation density
\begin{align}
    U_{Y}^{(N)} &= -g_{\sigma Y}\sigma+g_{\omega Y}\omega+\Sigma_{R},\label{equ:poten_depth_sigma} \\
    U_{Y}^{(Y)} &= -g_{\sigma Y}\sigma+g_{\omega Y}\omega -g_{\sigma^{*}Y}\sigma^{*}+g_{\phi Y}\phi+\Sigma_{R}. \label{equ:poten_depth_sigmas}
\end{align}
The $U_{Y}^{(Y)}$'s are much uncertain. For reducing the complexity of the discussion, we assume $U_{\Xi}^{(\Xi)}\simeq U_{\Lambda}^{(\Xi)}\simeq 2U_{\Lambda}^{(\Lambda)}\simeq 2U_{\Xi}^{(\Lambda)}\simeq -10$ MeV \citep{Takahashi2001_PRL87-212502,Lim2018_PRD97-023010,Schaffner1994_AoP235-35} and $g_{\sigma^{*}\Lambda}=g_{\sigma^{*}\Sigma}$. The $U_{\Lambda}^{(N)}$ is about 30 MeV \cite{Wang2010_PRC81-025801,Schaffner-Bielich2000_PRC62-034311}. The $U_{\Sigma}^{(N)}$ and $U_{\Xi}^{(N)}$ are still uncertain and thus we leave them as free parameters. The ranges of $U_{\Sigma}^{(N)}$ and $U_{\Xi}^{(N)}$ are taken as $-10~\mathrm{MeV}\leq U_{\Sigma}^{(N)}\leq 30~\mathrm{MeV}$ \cite{Providencia2019_FASS6-13} and $-23~\mathrm{MeV}\leq U_{\Xi}^{(N)}\leq -11~\mathrm{MeV}$ \cite{Harada2021_PRC103-024605} in this work. We hereinafter omit the superscripts of $U_{\Sigma}^{(N)}$ and $U_{\Xi}^{(N)}$ for simplicity.

In the present work, for every $L_{\mathrm{sym}}$, we change one of two groups parameters, i.e., ($z,~\alpha_{V}$) or ($U_{\Sigma},~U_{\Xi}$) while keeping the other one fixed, i.e., ($U_{\Sigma},~U_{\Xi}$) = (+30 MeV, $-$17 MeV) or ($z,~\alpha_{V}$)=($1/\sqrt{6},~1$), to change the hyperon coupling and investigate the effects of hyperon couplings on the hyperon superfluidity with different $L_{\mathrm{sym}}$.

\section{The hyperon superfluidity and the hypernuclear potentials}
\label{sec:discussion_hyperU}

\begin{figure}[tb]
\centering
\includegraphics[width=0.45\textwidth]{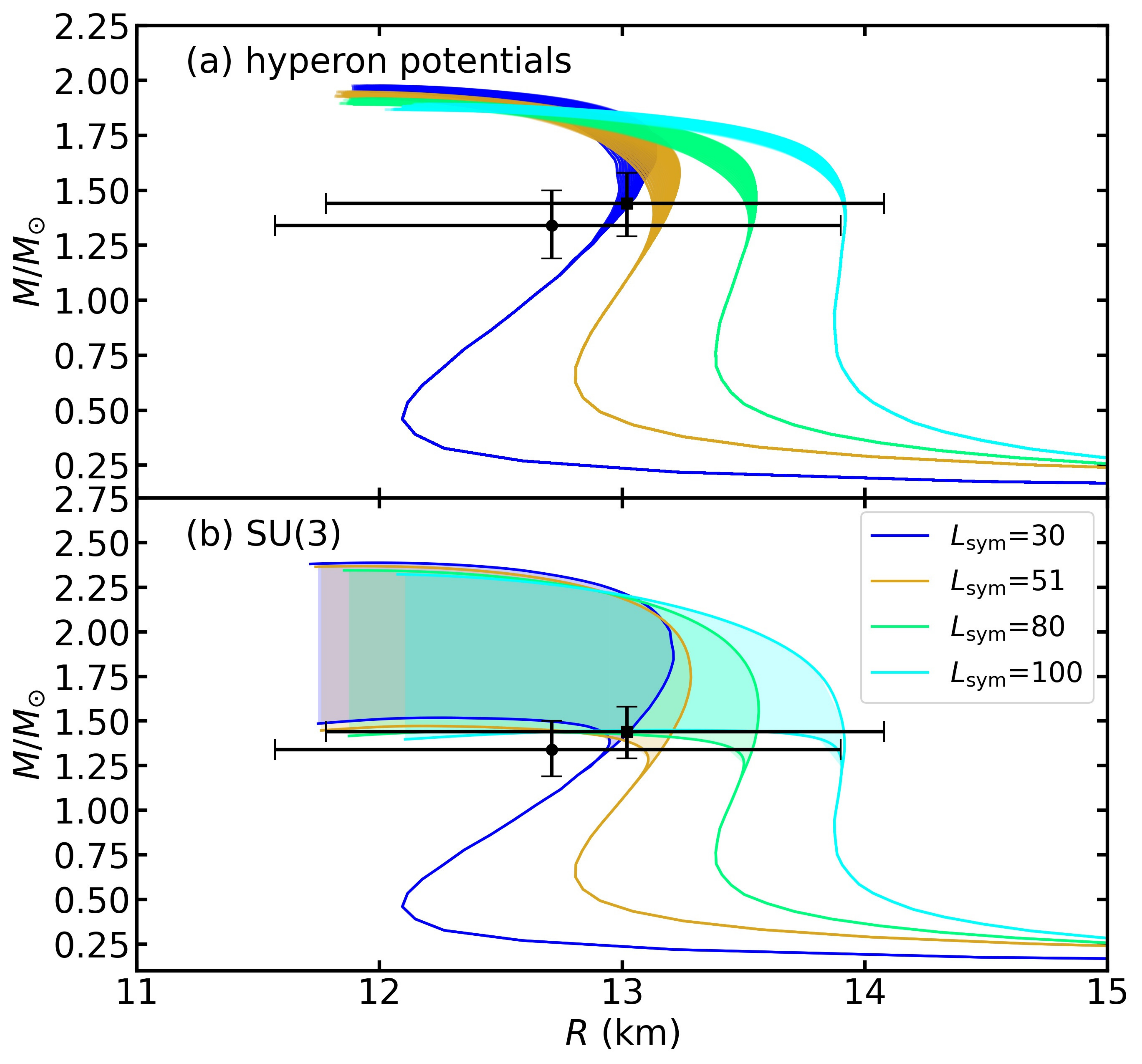}
\caption{The $M$-$R$ relations (a) obtained by varying ($U_{\Sigma},~U_{\Xi}$) and (b) under SU(3) symmetry. In the case of varying hyperon potentials, the all $M$-$R$ relations calculated with 625 combinations of ($U_{\Sigma},~U_{\Xi}$) for different $L_{\mathrm{sym}}$ are presented, the different combinations for ($U_{\Sigma},~U_{\Xi}$) are obtained by taking 25 points at equal intervals within the value range of $U_{\Sigma}$ and $U_{\Xi}$ respectively. Under SU(3) symmetry, the $M$-$R$ relations calculated by varying $\alpha_{V}$ but fixing $z=0.782476$ for different $L_{\mathrm{sym}}$ are illustrated. In panel (b), the top lines and bottom line stand for $\alpha_{V}=0.0$ and $\alpha_{V}=1.0$ respectively, the results between $\alpha_{V}=0.0$ and $\alpha_{V}=1.0$ are shown by shadows. The error bars are constraints from mass and radius of PSR J0030+0451 from NICER \cite{Riley2019_ApJL887-L21,Miller2019_ApJL887-L24}.
}
\label{fig:MR}
\end{figure}

In this section, we keep the parameters ($z,~\alpha_{V}$) fixed at their SU(6) values ($1/\sqrt{6},~1$). How the hyperon potentials affect the properties of NSs has been discussed before \cite{Weissenborn2012_NPA881-62,Providencia2019_FASS6-13}. When $U_{\Sigma}$ and $U_{\Xi}$ are considered at the same time, the maximum mass of NSs is insensitive to $U_{\Sigma}$ where $U_{\Sigma}>U_{\Xi}$ \cite{Weissenborn2012_NPA881-62}. In the parameter space ($U_{\Sigma},~U_{\Xi}$) in this work, $U_{\Sigma}$ is always larger than $U_{\Xi}$ and hence the maximum masses of NS for different $L_{\mathrm{sym}}$ mainly depend on $U_{\Xi}$. We take 25 points at equal intervals within the value range of $U_{\Sigma}$ and $U_{\Xi}$ respectively, and finally get 625 combinations of ($U_{\Sigma},U_{\Xi}$). The mass-radius ($M$-$R$) relations of NSs for all ($U_{\Sigma},U_{\Xi}$) and different $L_{\mathrm{sym}}$ are displayed in Fig. \ref{fig:MR}(a). We can see that $L_{\mathrm{sym}}$ have a distinct impact on the radius of NSs but the maximum masses of NSs are nearly independent of it ($\Delta M_{\mathrm{max}}<0.1M_{\odot}$). Although these radii differ significantly but all $M$-$R$ relations satisfy the constraints from NICER \cite{Riley2019_ApJL887-L21,Miller2019_ApJL887-L24}. The maximum masses are slightly lower than the observed most massive NSs, e.g., MSP J0740+6620 (2.08$^{+0.07}_{-0.07}M_{\odot}$ with 68.3\% credibility
interval)\cite{Fonseca2021_ApJL915-L12} and the estimation $M_{\mathrm{max}}>2.09M_{\odot}$ ($3\sigma$ confidence) of the minimum value for the maximum NS mass \cite{Romani2022_ApJL934-L17}. The modification of $K_{\mathrm{sat}}$ and $Q_{\mathrm{sat}}$ can increase the maximum mass, see Refs. \cite{Li2019_PRC100-015809,Fu2022_PLB834-137470}, we do not refer to these additional degrees of freedom in the present work. The changes in the hyperon potentials affect the hyperon couplings and therefor only influence the EoS above the onset density of the first appearing hyperon. As discussed in Ref. \cite{Weissenborn2012_NPA881-62}, hyperon potentials have slight impact on the mass and radius of NS. In our parameter space ($U_{\Sigma},~U_{\Xi}$), the variation of maximum mass induced by varying hyperon potentials is less than $0.03M_{\odot}$ and is below several uncertainties of astrophysical observation and analysis, e.g., $2.01M_{\odot}$ of PSR J0348+0432 with an uncertainty $0.04M_{\odot}$ \cite{Demorest2010_Nature467-1081}. It means that it is difficult to extract the information of hyperon potentials through astrophysical observations (e.g., mass and radius of NSs) and further explore the effects of these hypernuclear potentials on some microscopic physical processes of NSs (e.g., neutrino emission).

It is a failure to use global properties to resolve the hyperon potentials and interior compositions of neutron stars due to the approximate degeneracy of $M$-$R$ relations in parameter space ($U_{\Sigma},~U_{\Xi}$). In Ref. \cite{Providencia2019_FASS6-13}, the effects of $U_{\Sigma}$ on the interior composition (e.g., onset density and fraction of hyperon) of NSs have been discussed. The onset density and fraction of hyperon depend on $U_{\Sigma}$, especially for $\Sigma^{-}$ and $\Xi^{0,-}$. Here we briefly review the main result about $U_{\Sigma}$ and interior composition. With increasing of $U_{\Sigma}$, the onset density of $\Sigma^{-}$ is shifted to a higher value and the fraction of $\Sigma^{-}$ is suppressed. Those of $\Xi^{0,-}$ show the opposite trend compared to $\Sigma^-$. The pairing gap of hyperon is sensitive to interior composition and can potentially distinguish NSs with similar global properties but different $U_{\Sigma}$ and $U_{\Xi}$.

\begin{figure}[tb]
\centering
\includegraphics[width=0.45\textwidth]{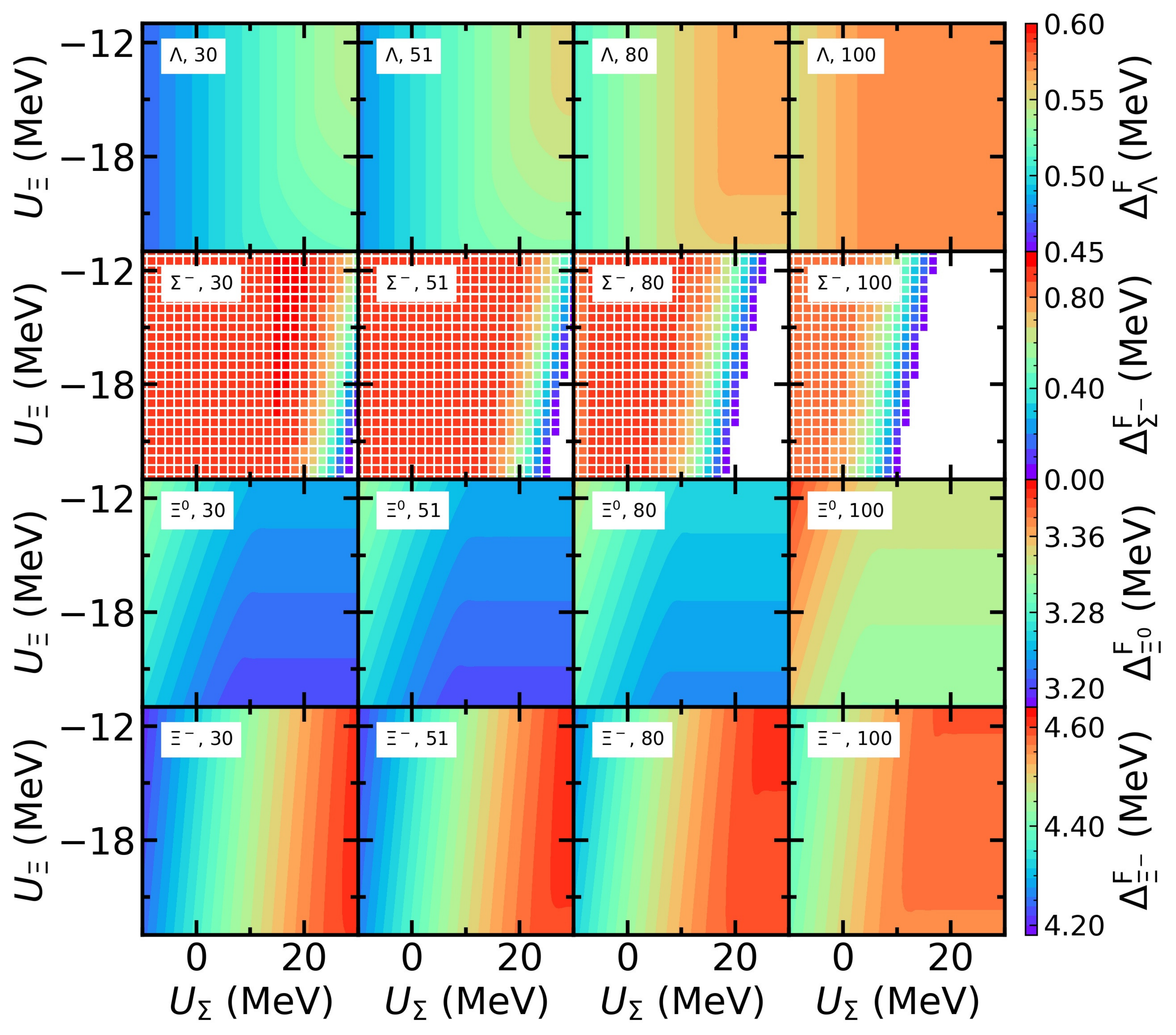}
\caption{The maximum pairing gaps of $\Lambda$, $\Sigma^-$, $\Xi^0$ and $\Xi^-$ inside NSs for different $L_{\mathrm{sym}}$ in the ($U_{\Sigma},U_{\Xi}$) parameter space. For $\Sigma^-$, there is no pairing at the zone with high $U_{\Sigma}$ and hence the heat map is used to display its result. The same row represent the results of the same hyperon and the same column represent those for the same $L_{\mathrm{sym}}$. The subgraph representing the maximum pairing gaps of hyperon $Y$ for $L_{\mathrm{sym}}$ is labeled by ($Y$,~$L_{\mathrm{sym}}$).
}
\label{fig:Gap_hyperU}
\end{figure}

\begin{figure}[tb]
\centering
\includegraphics[width=0.45\textwidth]{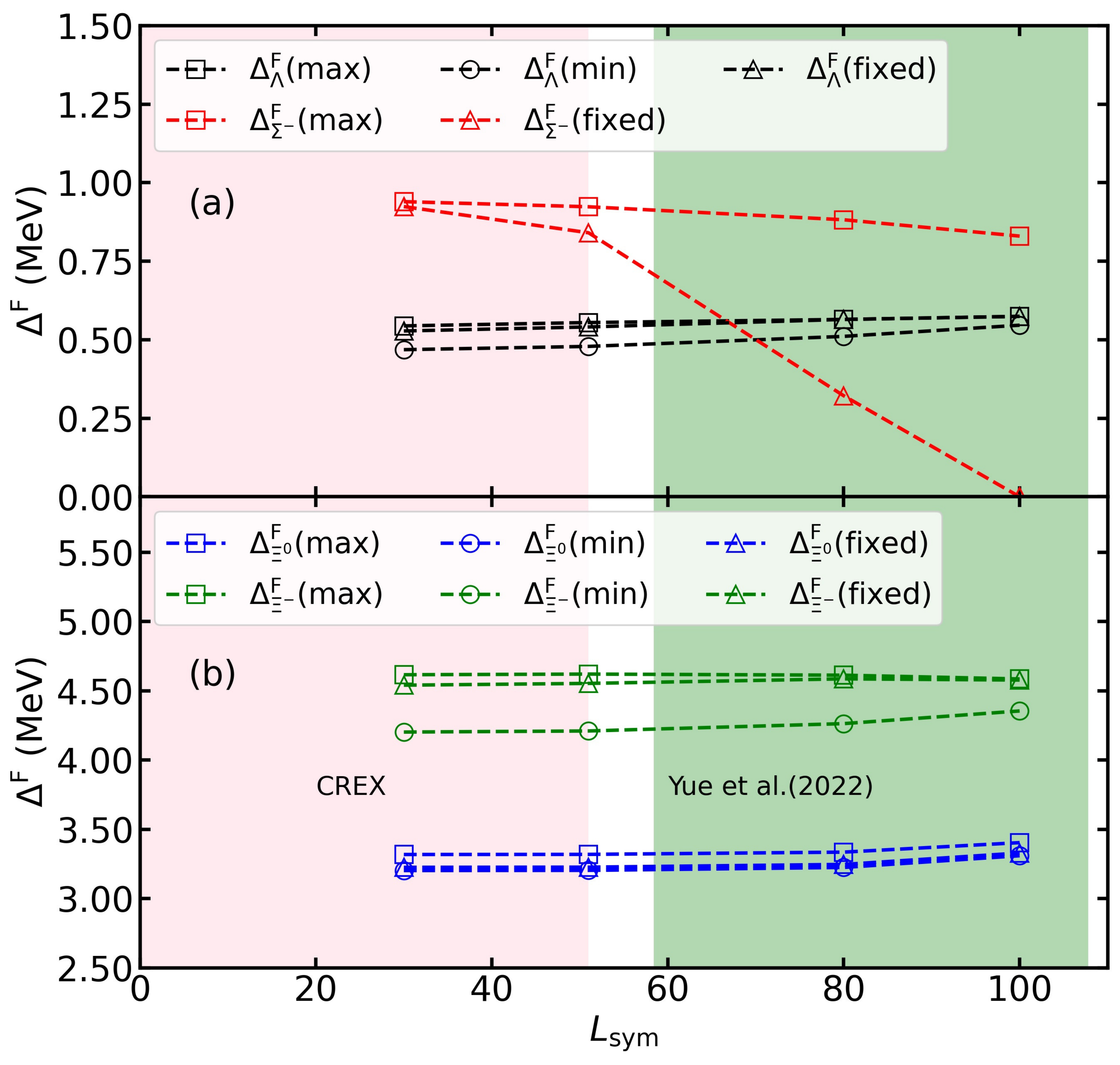}
\caption{The several pairing gaps of (a) $\Lambda$, $\Sigma^{-}$ and (b) $\Xi^{0,-}$ with the parameter space ($U_{\Sigma},~U_{\Xi}$) as a function of $L_{\mathrm{sym}}$ . The pink area is the constraint of $L_{\mathrm{sym}}$, i.e., $L_{\mathrm{sym}}=0$--51 MeV, from CREX \cite{Tagami2022_RP43-106037}. The constraint of $L_{\mathrm{sym}}$ with $83.1\pm24.7$ MeV from Ref. \cite{Yue2022_PRR4-L022054} is shown by green area. $\Delta_{\Lambda}^{\mathrm{F}}(\mathrm{max})$ is the maximum $\Delta_{\Lambda}^{\mathrm{F}}$, $\Delta_{\Lambda}^{\mathrm{F}}(\mathrm{min})$ is minimum one and $\Delta_{\Lambda}^{\mathrm{F}}(\mathrm{fixed})$ is the value of $\Delta_{\Lambda}^{\mathrm{F}}$ at $U_{\Sigma}=20$ MeV and $U_{\Xi}=-15$ MeV. There is no overlap between $L_{\mathrm{sym}}$ from CREX and Ref. \cite{Yue2022_PRR4-L022054}, the $L_{\mathrm{sym}}$ of original DD-ME2 lies this gap and those of other effective interactions cover the two constraints of $L_{\mathrm{sym}}$.
}
\label{fig:Gap_L_hyperU}
\end{figure}

In Fig. \ref{fig:Gap_hyperU}, the maximum pairing gaps $\Delta_{Y}^{\mathrm{F}}$ of $\Lambda$, $\Sigma^-$, $\Xi^0$ and $\Xi^-$ inside NSs for different $L_{\mathrm{sym}}$ within the ($U_{\Sigma},~U_{\Xi}$) parameter space are shown. For $\Lambda$, the pairing gaps are several tenths of a MeV and their variation $\Delta_{\Lambda}=\Delta_{\Lambda}^{\mathrm{F}}(\mathrm{max})-\Delta_{\Lambda}^{\mathrm{F}}(\mathrm{min})$ in the parameter space are smaller than 0.08 MeV, where $\Delta_{\Lambda}^{\mathrm{F}}(\mathrm{max})$ is the maximum $\Delta_{\Lambda}^{\mathrm{F}}$ and $\Delta_{\Lambda}^{\mathrm{F}}(\mathrm{min})$ is the minimum one. The variations are small and the dependence of $\Delta_{\Lambda}^{\mathrm{F}}$ to hyperon potentials are different from the maximum mass. Specifically, $\Delta_{\Lambda}^{\mathrm{F}}$ is slightly sensitive to $U_{\Sigma}$ but is insensitive to $U_{\Xi}$ at smaller $U_{\Sigma}$. As $U_{\Sigma}$ increases, the $\Lambda\Lambda$ pairing becomes stronger and $\Delta_{\Lambda}$ decreases. The trend can be visualized in Fig. \ref{fig:Gap_L_hyperU}(a). Due to the small variation of $\Delta_{\Lambda}^{\mathrm{F}}$, if only $L_{\mathrm{sym}}$ and hyperon potentials are considered, they have no significant impact on the $\Lambda\Lambda$ pairing gap and hence we can leave the $\Lambda\Lambda$ pairing as an invariant factor in NS cooling. Similar conclusions can be drawn for $\Xi^{0,-}$ with two main differences: firstly, both pairing gaps of $\Xi^{0,-}$ are an order of magnitude larger than that of $\Lambda$, see Fig. \ref{fig:Gap_L_hyperU}(a) and (b); secondly, the variation of $\Delta_{\Xi^{0}}^{\mathrm{F}}$ is significantly larger than that of $\Delta_{\Xi^{-}}^{\mathrm{F}}$, as shown in \ref{fig:Gap_L_hyperU}(b). The reason for the latter is that the onset density of $\Xi^{0}$ is considerably higher than those of $\Lambda$, $\Sigma^-$ and $\Xi^{-}$, resulting in the population of $\Xi^{0}$ being almost exclusively affected by $U_{\Xi}$. In fact, $\Delta_{\Xi^{0}}^{\mathrm{F}}$ is sensitive to $U_{\Sigma}$ at lower $U_{\Sigma}$ but depends on $U_{\Xi}$ at higher $U_{\Sigma}$. The variations for $\Lambda$ and $\Xi^{0,-}$ decreases as $L_{\mathrm{sym}}$ increases, which indicates that higher $L_{\mathrm{sym}}$ weakens the effects of hyperon potentials on hyperon superfluidity.

The effects of hyperon potentials on the pairing of $\Sigma^-$ are more interesting. From Fig. \ref{fig:Gap_hyperU}, there is no paired $\Sigma^-$ in the parameter space that falls within the blank area. We draw the heat map to illustrate the maximum pairing gaps of $\Sigma^-$ to avoid a smoothing problem. For $L_{\mathrm{sym}}=30$ MeV, the $\Sigma^-\Sigma^-$ pairing almost appears everywhere, the exception is the area where $U_{\Sigma}$ is close to its upper boundary and $U_{\Xi}$ approaches its minimum value. We can see that $\Delta_{\Sigma^-}^{\mathrm{F}}$ can reach about 1 MeV below a critical $U_{\Sigma}^{\mathrm{c}}=20$ MeV and the pairing gaps are relatively constant. As $U_{\Sigma}$ increases to around 51 MeV, the area where the paired $\Sigma^-$ appears becomes smaller, leaving a larger blank area in the lower right corner of the parameter space. The critical $U_{\Sigma}^{\mathrm{c}}$ also reduces to approximately 15 MeV. With the further increase of $L_{\mathrm{sym}}$, the blank area is extended to a quarter and a half of the parameter space for $L_{\mathrm{sym}}=80$ MeV and 100 MeV, $U_{\Sigma}^{\mathrm{c}}$ is reduced to 10 MeV and 0 MeV, respectively. The larger $L_{\mathrm{sym}}$ is unfavored to the $\Sigma^-\Sigma^-$ pairing.

The negative influence of the larger $L_{\mathrm{sym}}$ on the strength of $\Sigma^-\Sigma^-$ pairing can be discussed from global and local perspectives. From Fig. \ref{fig:Gap_hyperU} and Fig. \ref{fig:Gap_L_hyperU}(a), although the dependence of $\Delta_{\Sigma^-}^{\mathrm{F}}$ on ($U_{\Sigma},~U_{\Xi}$) is different quickly, the maximum of $\Delta_{\Sigma^-}^{\mathrm{F}}$ within the parameter space decreases slightly as $L_{\mathrm{sym}}$ increases. It means that the strong $\Sigma^-\Sigma^-$ pairing is always found below a critical $\Sigma$ hyperon potential $U_{\Sigma}^{\mathrm{c}}$. If $L_{\mathrm{sym}}$ is constrained well, the upper limit of $U_{\Sigma}$ (or $U_{\Sigma}^{\mathrm{c}}$) can be estimated by the verified of the existence of strong $\Sigma^-\Sigma^-$ inside NSs. From a local perspective, for a fixed ($U_{\Sigma},~U_{\Xi}$), the variation of $\Delta_{\Sigma^-}^{\mathrm{F}}(\mathrm{fixed})$ may be obvious. In Fig. \ref{fig:Gap_L_hyperU}(a), we can see that $\Delta_{\Sigma^-}^{\mathrm{F}}(\mathrm{fixed})$ at $U_{\Sigma}=20$ MeV and $U_{\Xi}=-15$ MeV drops to zero as $L_{\mathrm{sym}}$ increases. When $U_{\Sigma}$ and $U_{\Xi}$ are constrained, due to the sensitivity of $\Delta_{\Sigma^-}^{\mathrm{F}}$ to $L_{\mathrm{sym}}$, the value $L_{\mathrm{sym}}$ could be inferred by the determination of $\Delta_{\Sigma^-}^{\mathrm{F}}$.

\section{The hyperon superfluidity under the SU(3) symmetry}
\label{sec:discussion_SU3}

The massive $M\gtrsim 2.0M_{\odot}$ NSs featuring hyperons can be obtained if the value of $Q_{\mathrm{sym}}$ is large enough \cite{Li2019_PRC100-015809}. The modification of $Q_{\mathrm{sym}}$ is related to $NN$ interaction. In the $NY$ and $YY$ sectors, the repulsive interaction which increases the maximum mass of NS can be enhanced by relaxing SU(6) symmetry to the more general SU(3) symmetry. Indeed, the global properties and interior compositions of NSs within the SU(3) symmetry have been studied \cite{Weissenborn2012_PRC85-065802,Lim2018_PRD97-023010,Li2018_EPJA54-133,Fu2022_PLB834-137470}.
We review the main results again. The effects of $z$ or $\alpha_{V}$ on EoS and particle fraction of NSs are qualitatively similar \cite{Li2018_EPJA54-133}. As $z$ decreases from its upper bound, the onset densities of hyperons are pushed to the higher densities and hyperons may appear in a small core of NSs. It means that there have more nucleonic matter in NSs for smaller $z$, leading to stiffer EoS and larger maximum masses. The massive NSs, which mass could up to the mass $M\backsimeq2.5M_{\odot}$ of the secondary object of GW190814 \cite{Abbott2020_ApJL896-L44}, calculated with SU(3) symmetry are essentially nucleonic stars. It implies that hyperons are less important for the massive NSs. The problem is how the hyperon pairings inside NSs under the SU(3) symmetry.

We take ($U_{\Sigma},~U_{\Xi}$) as ($+30$ MeV,~$-17$ MeV) in this section. In Fig. \ref{fig:MR}(b), the $M$-$R$ relations calculated by varying $\alpha_{V}$ and fixing $z=0.782476$ for different $L_{\mathrm{sym}}$ are illustrated. The top and bottom lines represent the results for $\alpha_{V}=0.0$ and $\alpha_{V}=1.0$, respectively. Contrary to the results displayed in Fig. \ref{fig:MR}(a), the $M$-$R$ relations of NSs are distinctly different ($>0.85M_{\odot}$) within SU(3) symmetry. It provides a possibility to probe the interior composition through measurements of the global properties of NSs. The feasibility could be further improved by incorporating observations of NS cooling connected to hyperon pairings.

\begin{figure}[tb]
\centering
\includegraphics[width=0.45\textwidth]{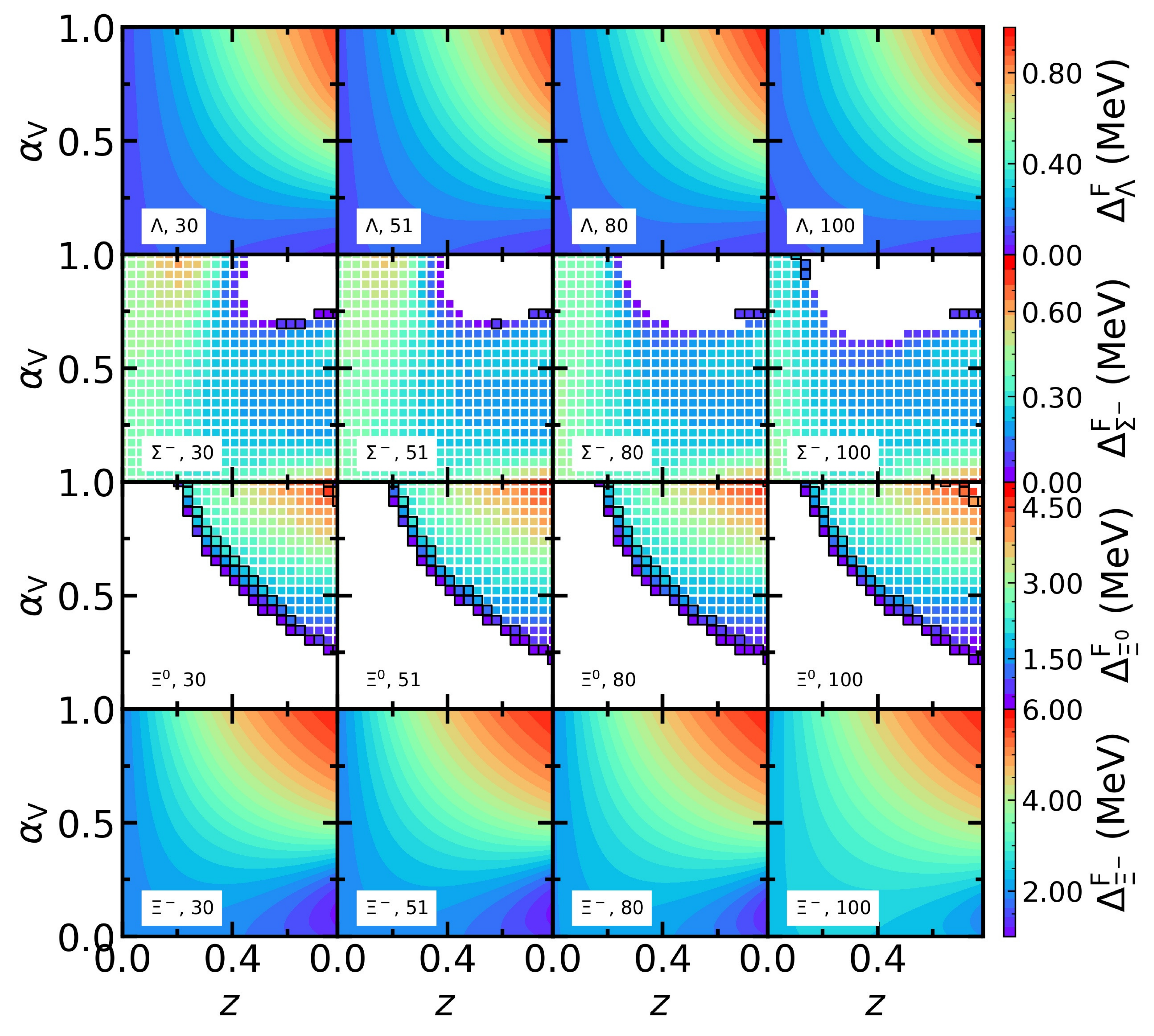}
\caption{The maximum pairing gaps of $\Lambda$, $\Sigma^-$, $\Xi^0$ and $\Xi^-$ inside NSs for different $L_{\mathrm{sym}}$ under SU(3) symmetry. For $\Sigma^-$ and $\Xi^0$, there is no pairing at the blank area and the heat map is used to display their results. The arrangement and label of the subgraphs are the same as those of Fig. \ref{fig:Gap_hyperU}. The black border in the results for $\Sigma^-$ and $\Xi^0$ means that the density corresponding to the maximum pairing gap exceed the central density of the most massive NSs and hence we use the pairing gap at the central density to replace the maximum pairing gap.
}
\label{fig:Gap_SU3}
\end{figure}

\begin{figure}[tb]
\centering
\includegraphics[width=0.45\textwidth]{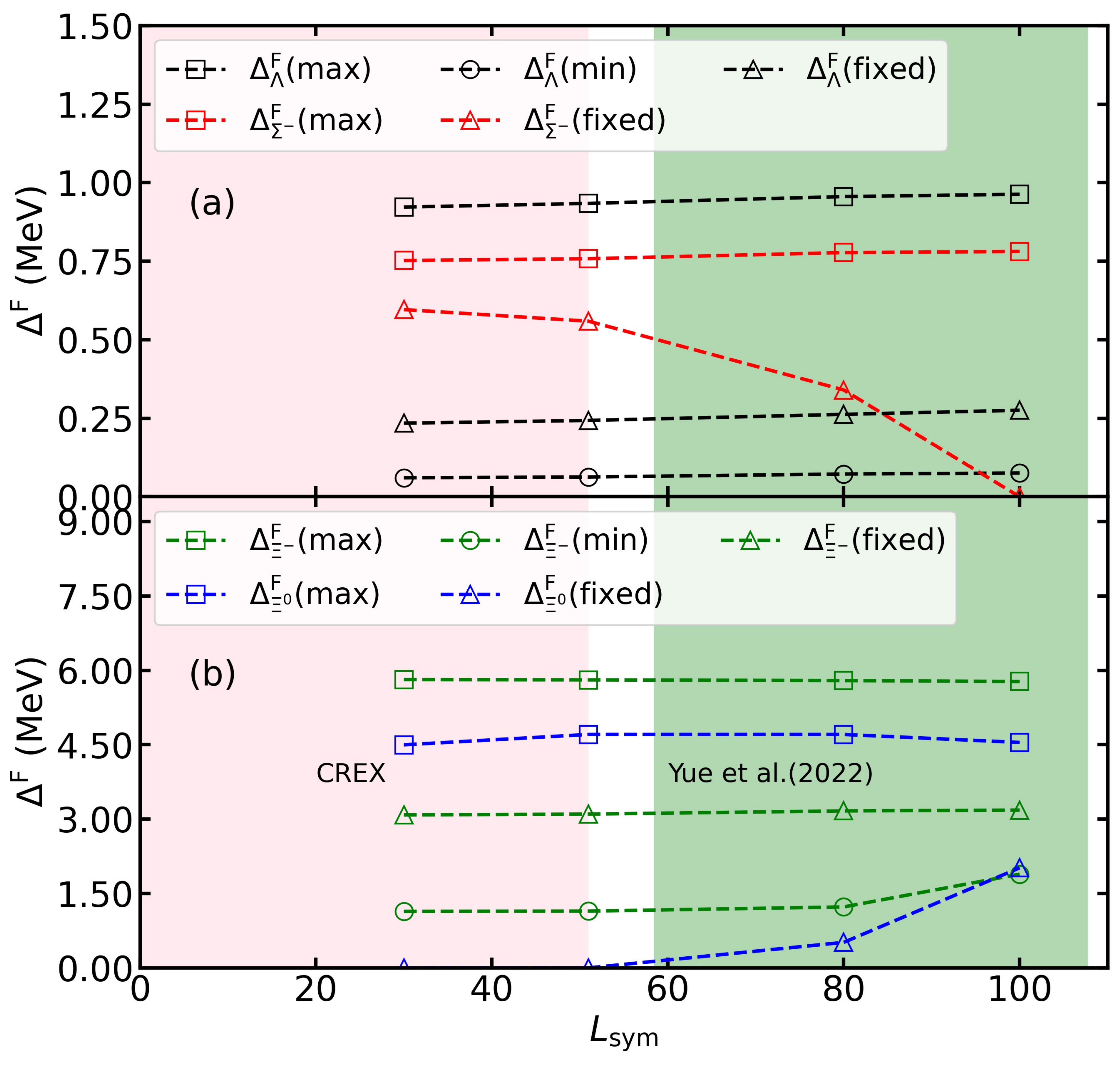}
\caption{The several pairing gaps of (a) $\Lambda$, $\Sigma^{-}$ and (b) $\Xi^{0,-}$ under SU(3) symmetry as a function of $L_{\mathrm{sym}}$. $\Delta_{\Lambda}^{\mathrm{F}}(\mathrm{fixed})$ is the value of $\Delta_{\Lambda}^{\mathrm{F}}$ at $z=0.170103$ and $\alpha_{V}=1.0$. The other definitions in figure are the same as Fig.\ref{fig:Gap_L_hyperU}.
}
\label{fig:Gap_L_SU3}
\end{figure}

In Fig. \ref{fig:Gap_SU3}, we demonstrate the maximum pairing gaps $\Delta_{Y}^{\mathrm{F}}$ of $\Lambda$, $\Sigma^-$, $\Xi^0$ and $\Xi^-$ inside NSs for different $L_{\mathrm{sym}}$ under SU(3) symmetry. For $\Lambda$ with a low baryon mass and the most depth potential and $\Xi^-$ with the negative charge and the moderate depth potential, they are the domain hyperons inside NSs and their onset densities are usually close to each other. From Fig. \ref{fig:Gap_SU3}, we can see that both the larger pairing gaps for $\Lambda$ and $\Xi^-$ locate in the upper right corner of SU(3) parameter space. Similar to EoS or NS mass, $z$ and $\alpha_{V}$ yield similar effects on the pairing gaps of $\Lambda$ and $\Xi^-$: the reducing of $z$ or $\alpha_{V}$ lead to the smaller pairing gaps. The massive NSs are unfavorable to $\Lambda$ and $\Xi^{-}$ superfluidity because their pairing gap and the maximum mass of NSs increase in opposite directions. The low fractions of $\Lambda$ and $\Xi^{-}$ reduce their state densities at the Fermi surface and further decrease the corresponding pairing gaps. The variations of pairing gaps are huge and their magnitudes are the same as those of the max pairing gaps, as shown in Fig. \ref{fig:Gap_L_SU3}. From Fig. \ref{fig:Gap_L_SU3}, we also find that $L_{\mathrm{sym}}$ has little impact on the pairing gap at every fixed ($z,~\alpha_{V}$) for $\Lambda$ and the maximum of pairing gaps of $\Xi^-$. Note that the larger $L_{\mathrm{sym}}$ ($>80$ MeV) leads to larger pairing gaps of $\Xi^-$ in the area where close to its minimum, like near extreme point.

For $\Xi^0$, the results are more complicated. In this work, the central densities of NSs we discussed are bounded below those of the theoretical most massive NSs ($\rho_{\mathrm{cM}}^{\mathrm{max}}$). For $\Xi^0$, the dependence of $\Delta_{\Xi^0}^{\mathrm{F}}$ on $z$ or $\alpha_V$ is similar, but $\Xi^0$ is not paired in the lower $z$ and lower $\alpha_V$ zone because the onset density is shifted to the density range above $\rho_{\mathrm{cM}}^{\mathrm{max}}$. Note that the density corresponding to the maximum pairing gap ($\rho_{\mathrm{cp}}^{\mathrm{max}}$) might surpass $\rho_{\mathrm{cM}}^{\mathrm{max}}$ and the pairing gap at $\rho_{\mathrm{cM}}^{\mathrm{max}}$ is plotted in Fig. \ref{fig:Gap_SU3} by using black border. Depending on $L_{\mathrm{sym}}$, the black border is also plotted in the strong $\Xi^0\Xi^0$ pairing zone, $\rho_{\mathrm{cM}}^{\mathrm{max}}$ is smaller and even below $\rho_{\mathrm{cp}}^{\mathrm{max}}$ because the EoS is softer as shown in Ref. \cite{Weissenborn2012_PRC85-065802}. As $L_{\mathrm{sym}}$ increases, the area where $\Xi^0\Xi^0$ pairing exists gets larger but changes little. From Fig. \ref{fig:Gap_L_SU3}(b), the maximum pairing gap of $\Xi^0$ under SU(3) symmetry varies little with $L_{\mathrm{sym}}$. However, the pairing gap near the boundary with and without $\Xi^0\Xi^0$ pairing may
alter significantly. For example, the pairing gap at $z=0.170103$ and $\alpha_V=1.0$ increases with increasing $L_{\mathrm{sym}}$. Due to both the existence of massive NSs and the vanishing $\Xi^0$ superfluidity in the region with lower $z$ and lower $\alpha_V$, it is unnecessary to consider $\Xi^0$ superfluidity in the cooling of massive NSs, the lower bound of NS mass depends on $L_{\mathrm{sym}}$.

The result of $\Sigma^-\Sigma^-$ pairing is also interesting as we discussed in the case of varying hyperon potential. Contrary to $\Lambda\Lambda$ or $\Xi^{0,-}\Xi^{0,-}$ pairing, with the decrease of $z$ or $\alpha_{V}$, the maximum pairing gap increases in general. Unlike $\Xi^0$, the area where $\Sigma^-\Sigma^-$ pairing exists becomes smaller as $L_{\mathrm{sym}}$ increases. A peak of $\Delta_{\Sigma^-}^{\mathrm{F}}$ in the parameter space can be found near the extreme point ($2/\sqrt{6},~1.0$), the other peak can be found near the SU(6) point if $L_{\mathrm{sym}}$ is small ($<80$ MeV). The pairing gaps around peaks are several tenths of a MeV which can be compared with the pairing gaps of $\Lambda$. From Fig. \ref{fig:Gap_L_SU3}(a), $\Delta_{\Sigma^-}^{\mathrm{F}}(\mathrm{max})$ is insensitive to $L_{\mathrm{sym}}$ but a  $\Delta_{\Sigma^-}^{\mathrm{F}}(\mathrm{fixed})$ which is fixed at $z=0.170103$ and $\alpha_V=1.0$ is sensitive to $L_{\mathrm{sym}}$. It means that, if the breaking of SU(6) is constrained well, the $L_{\mathrm{sym}}$ can be estimated by determining of $\Delta_{\Sigma^-}^{\mathrm{F}}(\mathrm{fixed})$ in NSs like we discussed in Sec. \ref{sec:discussion_hyperU}.

\section{Summary and perspective}
\label{sec:summary}

We explored the effects of hyperon couplings on the hyperon superfluidity inside NSs. Combining $NN$ effective interactions with different $L_{\mathrm{sym}}$, the coupling constants between the vector mesons and hyperons are obtained under the SU(3) symmetry; the coupling constants of the scalar mesons are adjusted by the hypernuclear potentials to be confirmed. A set of separable pairing forces of finite range is used in this work.

By fixing the ($z,~\alpha_{V}$) to the SU(6) values, we find that the variations of the pairing gaps of $\Lambda$ and $\Xi^{0,-}$ are relatively small but $\Delta_{\Sigma^-}^{\mathrm{F}}$ can range from 0.0 to around 1.0 MeV in the ($U_{\Sigma},~U_{\Xi}$) parameter space. $L_{\mathrm{sym}}$ has little effect on the maximum pairing gaps of all hyperons but could change significantly the pairing gap of $\Sigma^-$ at some fixed hyperon potentials.
Next we fixed ($U_{\Sigma},~U_{\Xi}$) as (+30 MeV, $-$17 MeV), the pairing gaps of $\Lambda$, $\Sigma^-$ and $\Xi^{0,-}$ differ significantly under SU(3) symmetry. The massive NS ($>2 M_{\odot}$) is unfavorable to the superfluidity of $\Lambda$ and $\Xi^{0,-}$ but is favorable to that of $\Sigma^-$. $L_{\mathrm{sym}}$ also has less effects on the pairing gaps of $\Lambda$ and $\Xi^-$ but obviously influences those of $\Sigma^-$ and $\Xi^0$ at some fixed additional SU(3) parameters.

The $\Sigma^-$ superfluidity could be a better pointer to detect the properties of NSs: both the cases of hyperon potentials and SU(3) symmetry, the pairing gap of $\Sigma^-$ changes significantly and hence is sensitive to ($U_{\Sigma},~U_{\Xi}$) and ($z,~\alpha_V$); when ($U_{\Sigma},~U_{\Xi}$) or ($z,~\alpha_V$) are constrained well, the pairing gap of $\Sigma^-$ could depend on $L_{\mathrm{sym}}$, the smaller $L_{\mathrm{sym}}$ is favored to $\Sigma^-$ superfluidity; the tendency of the pairing gap of $\Sigma^-$ to change with $z$ or $\alpha_V$ is the opposite of those of $\Lambda$ and $\Xi^{0,-}$. The effects of $\Sigma^{-}$ on the thermodynamic and dynamic evolutions of NSs could not be ignored.

Although plausible quantitative results are presented in this work, our main conclusions are more qualitative because many uncertainties are not included. In the $NN$ sector, except for $L_{\mathrm{sym}}$, $K_{\mathrm{sat}}$ and $Q_{\mathrm{sat}}$ even other higher order characteristic coefficients also have significant effects on the EoS beyond twice the saturation density. In the $NY$ and $YY$ sectors, we keep $U_{Y}^{(Y)}$ fixed but the properties of NSs are also affected by them, see Ref. \cite{Li2018_EPJA54-133}. Uncertainties in pairing forces for various hyperons are also disregarded. Fortunately, the strength of the pairing force can be decomposed from the integral in Eq. \ref{equ:gap_equ} so that the relation between the pairing gap and the strength of the pairing gap is monotonic, the main conclusions obtained in this work still hold for different pairing strengths. We apply the standard BCS approximation in this work and neglect the many-body correlations, e.g., short- and long-range correlations. In Ref. \cite{Ding2016_PRC94-025802}, the authors have confirmed that the short-range correlations suppress the pairing gap of nucleon. The theoretical framework that accounts for many-body correlations is more realistic. However, our understanding of the correlations referring to hyperon-hyperon or nucleon-hyperon is lacking, we discard the treatment of many-body correlations to avoid introducing additional uncertainties. More quantitative results are expected to be obtained in our future works after some additional factors we mentioned above are well-constrained.

\nolinenumbers
\section*{Acknowledgements}
Valuable effective interaction parameters from Jia-Jie Li are gratefully acknowledged. This work has been supported by the National Natural Science Foundation of China (Grants No. 12070131001, No. 12047503, No. 12175151, No. 12375118, and No. 11961141004),
the Strategic Priority Research Program of Chinese Academy of Sciences (Grant No. XDB34010000), and the IAEA Coordinated Research Project ``F41033''. The results described in this paper are obtained on the High-performance Computing Cluster of ITP-CAS and the ScGrid of the Supercomputing Center, Computer Network Information Center of Chinese Academy of Sciences.

\bibliographystyle{elsarticle-num}
\bibliography{zhtu}

\end{document}